\documentclass[%
 reprint,
 amsmath,amssymb,
 aps,
]{revtex4-1}
\usepackage{amsthm}
\usepackage{amssymb}
\usepackage{braket}
\usepackage{graphicx}
\usepackage{tikz}
\usetikzlibrary{arrows}
\usepackage{subfig}
\usepackage{pgfplots}
\usepackage{xifthen}
\usepackage{float}
\usepackage{amsthm}
\usepackage{amssymb}
\usepackage{braket}

\newtheorem{thm}{Theorem}
\newtheorem{lem}[thm]{Lemma}
\newtheorem{claim}{Claim}
\newtheorem{define}{Definition}
\usepackage{graphicx}% Include figure files
\usepackage{dcolumn}% Align table columns on decimal point
\usepackage{bm}% bold math
\begin{document}

%\preprint{APS/123-QED}

\title{Continuous Limit of Discrete Quantum Walks}

\author{Dheeraj M N}
\email{ee11b091@ee.iitm.ac.in}
\affiliation{Department of Electrical Engineering, IIT Madras, Chennai, Tamil Nadu, India}

%\collaboration{MUSO Collaboration}%\noaffiliation

\author{Todd A. Brun}
\email{tbrun@usc.edu}
\affiliation{
Communication Sciences Institute, University of Southern California,
Los Angeles, California, USA
}%

\date{\today}

\begin{abstract}
Quantum walks can be defined in two quite distinct ways: discrete-time and continuous-time quantum walks (DTQWs and CTQWs).  For classical random walks, there is a natural sense in which continuous-time walks are a limit of discrete-time walks. Quantum mechanically, in the discrete-time case, an additional ``coin space'' must be appended for the walk to have nontrivial time evolution. Continuous-time quantum walks, however,  have no such constraints. This means that there is no completely straightforward way to treat a CTQW as a limit of DTQW, as can be done in the classical case.  Various approaches to this problem have been taken in the past.  We give a construction for walks on $d$-regular, $d$-colorable graphs when the coin flip operator is Hermitian:  from a standard DTQW we construct a family of discrete-time walks with a well-defined continuous-time limit on a related graph. One can think of this limit as a {\it coined} continuous-time walk.  We show that these CTQWs share some properties with coined DTQWs.  In particular, we look at spatial search by a DTQW over the 2-D torus (a grid with periodic boundary conditions) of size $\sqrt{N}\times\sqrt{N}$, where it was shown  \nocite{AAmbainis08} that a coined DTQW can search in time $O(\sqrt{N}\log{N})$, but a standard CTQW \nocite{Childs2004} takes $\Omega(N)$ time to search for a marked element. The continuous limit of the DTQW search over the 2-D torus exhibits the $O(\sqrt{N}\log{N})$ scaling, like the coined walk it is derived from. We also look at the effects of graph symmetry  on the limiting walk, and show that the properties are similar to those of the DTQW as shown in \cite{HariKrovi2007}.
\end{abstract}

%\pacs{Valid PACS appear here}% PACS, the Physics and Astronomy
                             % Classification Scheme.
%\keywords{Suggested keywords}%Use showkeys class option if keyword
                              %display desired
\maketitle

%\tableofcontents

\section{\label{sec:level1}Introduction}

Quantum walks are unitary analogues of classical random walks, and have many applications in quantum computing, as well as being interesting objects in their own right. Quantum walks are defined separately for discrete time (DTQW) \cite{AharonovY1993,Meyer1998,AharonovD2001,Ambainis2001,Nayak2000,Kempe2003} and continuous time (CTQW) cases \cite{FarhiGutmann1998,Childs2002,ChildsCleve2003}.

Algorithms based on classical random walks can solve a variety of classical computational problems efficiently, as shown in \cite{MotwaniRA1995}. The quantum analogues also have a wide variety of applications in quantum computation. They can be used to solve the element distinctness problem  \cite{Ambainis2007}. A QW-based search algorithm over the hypercube \cite{Shenvi2003} performs as efficiently as Grover's algorithm (\cite{Grover1996}), i.e, in time $O(\sqrt{N})$ for a database of size $N$. More applications of DTQWs are described in \cite{AAmbainisQwalk2004}.  CTQWs can solve the (albeit somewhat artificial) ``glued-trees'' problem exponentially faster than the best classical algorithm \cite{ChildsCleve2003}.  They give a polynomial speed-up in evaluating the NAND tree \cite{Farhi07aquantum}, which has been generalized to evaluating any Boolean formula \cite{Reichardt2012}.

While algorithms have been found based on both DTQWs and CTQWs, these walks cannot necessarily be used interchangeably, unlike the classical case.  To maintain both unitarity and nontrivial dynamics in a DTQW, the state space is expanded to have a ``coin space.''  Such walks are often called ``coined'' walks. This problems does not arise in CTQWs.  So CTQWs and DTQWs on the same graph have state spaces with different dimensions.  This makes it difficult to define a sequence of DTQWs having a CTQW as a limit, as can be done with classical random walks.

There have been a number of previous studies of this problem.  In \cite{Patel2005} a discrete-time walk without a coin is defined by alternating unitaries. In \cite{Strauch2006}, a correspondence is shown between the limiting behavior of CTQWs and DTQWs on the infinite line and the 3D square lattice are shown, and this is extended to general graphs in \cite{DAlessandro2009}.  In \cite{Molfetta2011}, a continuous-time limit is found for a limited class of one-dimensional quantum walks.

In \cite{Childs2010} Childs gives a discussion of the relationship between CTQWs and DTQWs, and presents a construction to discretize in the time domain, transforming a CTQW to a DTQW, and also shows how to recover the CTQW from the discretized DTQW by isometry mapping.  In this paper, we are mainly interested in starting from discrete time and producing a continuous time limit.  Moreover, our approach differs from previous work, in that the resulting CTQW is a {\it coined} continuous-time walk.  Differences between coined DTQWs and uncoined CTQWs have been noted in the past; we will see that the behavior of our coined CTQWs is more similar to that of the coined DTQWs than to standard CTQWs.

We consider one particular algorithm that exhibits a difference between CTQW and DTQW implementations.  The DTQW search on a 2D grid of size $\sqrt{N}\times\sqrt{N}$ with periodic boundary conditions was studied in \cite{AAmbainis08}, where it was shown that a probabilistic search can be done in $O(\sqrt{N}\log{N})$. But in \cite{Childs2004} it was shown that any CTQW search algorithm takes time of $\Omega(N)$ on the same graph. We show that by taking the continuous time limit of the DTQW search from \cite{AAmbainis08}, the search scaling of $O(\sqrt{N}\log{N})$ is recovered.

We also look at another property of DTQWs in the continuous time limit.  In \cite{HariKrovi2007}, it is shown that a DTQW on a graph with certain symmetries can be reduced to a walk on a smaller graph (the ``quotient graph'') for certain unitary time operators and initial conditions. There are many interesting consequences of such symmetry; for example, infinite hitting times for certain initial conditions as shown in \cite{HariKrovi2006}. In this paper, we show that the limiting CTQW inherits the symmetries from the DTQW.

In the next section we give the standard definitions of DTQWs and CTQWs, and compare their definitions to the continuous-time limit of a classical random walk.  In Sec.~III, we present the construction of a family of DTQWs on a graph with a well-defined continuous-time limit.  In Sec.~IV we apply this construction to the discrete-time walk-based search algorithm on the torus, and show that the continuous time limit of this walk exhibits the same scaling with the grid area $N$.  In Sec.~V we look at the effects of graph symmetry, and show that if a standard DTQW has a reduction to a quotient graph due to graph symmetry, the family of DTQWs also has a reduction to the quotient graph, including the continuous-time limit.  In Sec.~VI we conclude.

%%%%%%%%%%%%%%%%%%%%%%%%%%%%%%%%%%%%%%%%%%%%%%

\section{Definitions of quantum walks}

\subsection{Discrete- and continuous-time quantum walks}

We now define DTQWs and CTQWs. These definitions are taken from \cite{HariKrovi2007} and apply to $d$-regular graphs, but, the definitions can be extended to irregular graphs.   Let $G$ be a  $d$-regular graph on which the walk is defined. The Hilbert space of the DTQW is $\mathcal{H}^c\otimes\mathcal{H}^p$. In the position space $\mathcal{H}^p$, a basis vector $\ket{v}$ is associated with each of the vertices $v$, and in the coin space $\mathcal{H}^c$, a basis vector $\ket{i}$ is associated with each of the edges emanating from each vertex of G; $i$ is a label of the different directions one can walk.  The basis vectors of the coin and vertex states together are $\{\ket{i,v} \equiv \ket{i}\otimes\ket{v} \}$.

\begin{define}
The time evolution of a state vector in a \emph{DTQW} is $\ket{\Psi_{n+1}}=U\ket{\Psi_{n}}$, where $U=SF$, where
\begin{equation}
S = \sum_v\sum_i \ket{j(i,v),v(i)}\bra{i,v}
\label{shift_op}
\end{equation}
and
\begin{equation}
F = C \otimes I .
\end{equation}
\end{define}
In this definition, $S$ is the {\it shift} operator and $F$ is the {\it coin-flip} operator.  These are both unitary. Here, $v$ is any vertex of $G$; $v(i)$ is the vertex connected to $v$ along the direction $i$; and $j(i,v)$ is the direction by which $v(i)$  is connected back to $v$.  If $G$ is $d$-colorable, then we can always choose $j(v,i) = i$, so walking the the same direction twice takes one back to the vertex where one started.  We will assume that later in our construction.  In this case, $S$ is not only unitary but also Hermitian.

Strictly speaking, $F$ needn't have the tensor product structure $C\otimes I$. It could be of the form $F = \sum_i C_i\otimes P_i $ where $P_i$ are projection operators such that $\sum_i P_i = I $ (identity over $\mathcal{H}^p$) and the $C_i$ are unitary.  This would allow the coin to differ at different parts of the graph.  This type of coin is used in the search algorithm presented in this paper to ``mark" the node to be found.
\begin{define}
A \emph{CTQW} over $G$ is defined by the unitary tranformation $U(t)= e^{-iHt}$ and the state vector at any time $t$ is $\ket{\Psi(t)} = U(t)\ket{\Psi(0)}$. Here, $H=H^\dagger$ is a Hermitian operator such that for vertices $i \neq j$,
$$
   H_{ij}  \left\{
     \begin{array}{lr}
      \neq 0  $ if i and j share an edge,$\\
       = 0  $ otherwise,$
     \end{array}
   \right.
$$
and $H_{ii} \in \mathbb{R}$.
\end{define}
There is a canonical choice of Hamiltonian $H$ that one might call a {\it standard} CTQW.  Let $A=[A_{ij}]$ be the adjacency matrix of the graph $G$, so $a_{ij} = 1$ if $i$ and $j$ are connected by an edge and $a_{ij}=0$ otherwise; then for $i\ne j$, $H_{ij} = \kappa a_{ij}$, and $H_{ii} = \kappa d_i$, where $\kappa$ is an energy scale (or rate) and $d_i$ is the degree of vertex $i$.

The CTQWs defined in this paper are not of this standard form; but they do satisfy the broader definition above.

\subsection{Comparison with classical random walks}

Discrete time classical random walks, which are a special case of Markov chains, admit a continuous time limit. These are defined by linear difference equations with probabilities represented as vectors. The limiting process is well established and is found in most discussions of classical random walks (for example \cite{norris1998markov}).

Let a classical random walk be defined over a graph $G$. Let $p_n$ be a vector whose $i$th entry is the probability of being at the vertex $i$ at time step $n$. The time evolution is given by
\begin{equation}
p_{n+1} = Mp_{n} ,
\end{equation}
where $M = [m_{ij}]$ is a stochastic matrix.  The probability $m_{ji}$ to go from vertex $i$ to vertex $j$ is zero unless an edge connects $i$ to $j$.  In a standard undirected random walk on $G$, from a vertex $i$ there is an equal probability to walk along any of the edges connected to $i$, so $m_{ji} = 1/d_i$ where $d_i$ is the degree of vertex $i$.

As seen in \cite{Childs2010}, we can replace $M$ by $\epsilon M+ (1-\epsilon)I $ to obtain a family of walks parametrized by $\epsilon$.  The standard discrete-time random walk corresponds to $\epsilon=1$.  Taking the limit $\epsilon \rightarrow 0$ while defining the time to be $t_n = n\epsilon$, we obtain a  differential equation for the probability vector $p$:
\begin{equation}
\frac{dp(t)}{dt} = (M-I)p(t) .
\end{equation}
This equation gives the continuous-time limit of the discrete time classical random walk.

It is clear that both the discrete-time random walk and its continuous-time limit have the same number of states in the Markov chain. But, in the quantum case, the continuous time dynamics is defined through the Schr\"odinger equation
\begin{equation}
\frac{d\ket{\Psi(t)}}{dt} = -iH\ket{\Psi(t)},
\end{equation}
where the state space is of the same dimension as the number of vertices in $G$. But, as described above, the state space of the DTQW also includes the ``coin space," and thus the dimensions of the  state spaces of a DTQW and a CTQW over the same graph $G$ are different. We will overcome this difficulty by retaining the coin space in the continuous limit of the DTQW.  We can then think of this as either a CTQW over a different (but related) graph $G'$, or as a {\it coined} continuous-time walk.

\section{\label{sec:limit}Continuous time limit}

In this section, we show how to construct a family of quantum walks, starting from a DTQW defined on a regular, $d$-colorable, undirected graph of degree $d$ with a Hermitian coin flip operator $F$.  This family of walks is parametrized by a real number $s>0$, and has as a limit as $s \rightarrow0$ a CTQW over a different but related graph.  This walk can be considered a continuous-time coined walk.  This construction is  illustrated with an example, and a few properties of the continuous time limit are derived.

Let $G$ be an undirected graph with a DTQW defined by the shift operator $S$ and coin flip operator $F$.  $S$ is defined by Eq.~(\ref{shift_op}), with $j(i,v)=i$ (which can always be done if $G$ is regular and $d$-colorable). By the definition of $S$, it is Hermitian. We assume also that $F$ is Hermitian; many widely studied coin operators (e.g., the Hadamard and Grover coins) satisfy this assumption.  In this standard form of a DTQW, the shift operator $S$ is a Hermitian permutation matrix of order $2$ and the coin flip operator $F$ maps a basis state of the coin space associated with a vertex $v$ to a superposition of such basis states.

\begin{define}
A unitary transformation $U$ acting on the state space of a graph $G$ is a \emph{local transformation} if it maps any vector associated with a vertex $v$ to a superposition of vectors associated with $v$ and those associated with vertices sharing an edge with $v$.
\end{define}

\subsection{Family of DTQWs and limit}

Constructing the family of DTQWs is based on a simple property of operators that are both Hermitian and unitary.  If a finite dimensional operator $A$ is Hermitian and unitary, then $A^2=I$, and
\begin{equation}
e^{-i\frac{\pi}{2}(A-I)} = A .
\end{equation}
By assumption, both $S$ and $F$ are Hermitian. Define a family of step operators
\begin{equation}
U(s) = e^{-i\frac{\pi}{2}s(S-I)}e^{-i\frac{\pi}{2}s(F-I)} .
\end{equation}
where $s$ is a parameter $s \in [0,1]$.
\begin{lem}
$U(s)$ is a local transformation for all $s$, and $U(1) = SF$
\label{lem:local}
\begin{proof}
Since $S$ and $F$ are both Hermitian and unitary, $S^2=F^2=I$.  This implies that
\[
e^{-i\frac{\pi}{2}s(S-I)} = e^{i\frac{\pi}{2}s}(\cos(\pi s/2) I - i\sin(\pi s/2)S) ,
\]
and similarly for $F$.  Hence, $U(s) =  e^{i\pi s} ( \cos^2(\pi s/2) I - i\cos(\pi s/2)\sin(\pi s/2)(S+F) - \sin^2(\pi s/2) SF)$. Since $I$, $S$, $F$ and $SF$ are all local transformations over $G$, $U(s)$ is also a local transformation over $G$.  If we take $s=1$ then $\cos(\pi s/2) = 0$ and $e^{i\pi s}=-1$, so we get $U(1)=SF$.
\end{proof}
\end{lem}
Let the family of local transformations $U(s)$ be called $\mathcal{F}$. $SF \in \mathcal{F}$.  As $s \rightarrow 0$ , $U(s) \rightarrow I - i\frac{\pi}{2}s(S+F-2I)+O(s^2)$.  This shows that as $s\rightarrow 0$, the local transformations in $\mathcal{F}$ behave like the CTQW over another graph $G^{\prime}$ with Hamiltonian $\mathcal{H} = S+F-2I$ (up to a time scaling of $\frac{\pi}{2}$). The connectivity of $G^{\prime}$ is described by the entries of $\mathcal{H}$ as seen in the definition of CTQW. Hence, this CTQW can be seen as the limit of the family of local transformations $\mathcal{F}$ as $s \rightarrow 0$. The continuous-time limit defined by $\mathcal{H} = S+F-2I$ is equivalent to that defined by $\mathcal{H} = S+F$ up to a global phase.

\subsection{Relationship between the original and new graph:  the coined continuous-time walk}

Henceforth in the paper, we denote with a prime the graph over which the limiting CTQW of a DTQW is defined; the original graph is unprimed. The number of vertices in the limiting graph $G^{\prime}$ is the same as the dimension of the state space of $G$.  Hence, we can index the states associated with $G^{\prime}$ with the same labels as the states associated with $G$.  Each vertex subspace of the original walk is mapped onto a collection of vertices in $G'$.  The edges among these vertices is given by the coin flip operator $F$, and each of them is connected to a neighboring vertex of the original graph $G$.  We can group these collections of vertices together, and consider this limiting case to be a coined CTQW.

Note that if we allow self loops, then there are multiple possible graphs $G'$ on which $\mathcal{H}$ defines a CTQW. We therefore consider only the one $G'$ which has no self loops.

\paragraph*{Example}  Consider the graph $G$ to be the square shown in Fig.~\ref{fig:g}.  Because this is a cycle ($d=2$) with an even number of vertices ($n=4$), we can define a DTQW with a coin of dimension 2, and the graph $G$ is 2-colorable.  The coin flip operator $F$ could be any $2\times2$ matrix that is both Hermitian and unitary; the Hadamard is a commonly used choice. The CTQW is defined on the graph $G^{\prime}$ as shown in Fig.~\ref{fig:gdash}.  Each vertex of the original graph $G$ is mapped to two vertices of $G'$, one for each coin state of each vertex; edges between coin states of the same vertex represent ``coin flip'' transitions, while edges between coin states of different vertices represent ``shift'' transitions.

\begin{figure}[ht]
\centering
\begin{tikzpicture}[-,>=stealth',shorten >=1pt,auto,node distance=3cm,
  thick,main node/.style={circle,draw,font=\sffamily\small\bfseries}]
\node[main node] (0){} ;
\node[main node] (1) [right of=0]{};
\node[main node] (2) [below of=0]{};
\node[main node] (3) [right of=2]{};
\path[every node/.style={font=\sffamily\small}]
(0) edge node {} (2)
(0) edge node {} (1)
(2) edge node {} (3)
(3) edge node {} (1);
%(0) edge node [bend right] {c} (7)
%(7) edge node [bend right] {b} (4)
%(6) edge node [bend right] {c} (1)
%(1) edge node [bend right] {b} (2)
%(3) edge[bend left] node [bend left] {b} (0)
%(4) edge[bend left] node [bend left] {c} (3)
%(2) edge[bend left] node [bend left] {c} (5)
%(5) edge[bend left] node [bend left] {b} (6)
%(1) edge [bend right] node [bend right] {d} (7)
%(7) edge [bend right] node [bend right] {d} (1)
%(2) edge[bend right]node [bend right] {a} (0)
%(6) edge[bend right] node [bend right] {a} (4)
%(4) edge[bend right] node [bend right] {a} (6);
\end{tikzpicture}
\caption{graph $G$}
\label{fig:g}
\end{figure}

\begin{figure}[ht]
\centering
\begin{tikzpicture}[-,>=stealth',shorten >=1pt,auto,node distance=2cm,
  thick,main node/.style={circle,draw,font=\sffamily\small\bfseries}]
\node[main node] (0){} ;
\node[main node] (1) [right of=0]{};
\node[main node] (2) [below of=0]{};
\node[main node] (3) [right of=2]{};
\node[main node] (4) [above left of = 0]{} ;
\node[main node] (5) [above right of=1]{};
\node[main node] (6) [above left of=2]{};
\node[main node] (7) [above right of=3]{};
\path[every node/.style={font=\sffamily\small}]
(0) edge node {} (4)
(1) edge node {} (5)
(4) edge node {} (6)
(7) edge node {} (5)
(2) edge node {} (6)
(3) edge node {} (7)
(0) edge node {} (1)
(2) edge node {} (3);
\end{tikzpicture}
\caption{graph $G^{\prime}$}
\label{fig:gdash}
\end{figure}

\subsection{Properties of continuous-time limit}

The evolution of the DTQW is the same if the coin $F$ is replaced with $-F$ except for a time-dependent global phase.  However, the family of maps may differ. The requirement for this transformation will become apparent in the sections to follow, during the analysis of the search algorithm.
\begin{lem}
\label{lem:exp}
If $\ket{j}$ is an eigenvector of $SF$ with eigenvalue $e^{i\phi_j}$, then $\ket{j}$ is also an eigenvector of $(S-F)^2$ with eigenvalue $4\sin^2{\frac{\phi_j}{2}}$. So if $SF = \sum_j e^{i\phi_j}\ket{j}\bra{j}$ then
\[
(S-F)^2 = \sum_j 4\sin^2{\frac{\phi_j}{2}}\ket{j}\bra{j} .
\]
Define $\lambda_j = 2\sin{\frac{\phi_j}{2}}$.  Then
\[
e^{-i(S-F)t} = \sum_j \left( \cos{\lambda_jt}\ket{j}\bra{j} -i\frac{\sin{\lambda_jt}}{\lambda_j}\ket{j}\bra{j}(S-F) \right) .
\]
\end{lem}
When $\phi_j$ is sufficiently small,  $\lambda_j \approx \phi_j$. Roughly speaking, this means that the limiting walk defined by the Hamiltonian $\mathcal{H}=S-F$ approximates the DTQW pretty well, as far as time evolution in the state space is concerned.

%%%%%%%%%%%%%%%%%%%search algorithm%%%%%%%%%%%%%%%%%%%%%%%%%%
\section{Search algorithm}

In the search algorithm as described in \cite{AAmbainis08}, $S$ is the shift operator over a 2D $\sqrt{N}\times\sqrt{N}$ grid with periodic boundary conditions, and the coin flip operator is
\begin{equation}
F = C_0\otimes I - (C_0-C_1)\otimes\ket{x}\bra{x},
\end{equation}
where
\begin{equation}
C_0 = 2\ket{S_c}\bra{S_c}-I,\ \ \ 
\ket{S_c} = \frac{1}{\sqrt{d}}\sum_i\ket{i},\ \ \ 
C_1 = -I .
\end{equation}
$C_0$ is called the {\it Grover coin}.  The state $\ket{S_c}$ is the uniform superposition of all coin states, and $d$ is the dimension of coin space.  The state $\ket{x}$ is the marked vertex which is to be found. The idea behind the algorithm is that we use the coin flip $C_0$ on all vertices other than $x$, and use the coin flip $C_1$ on $x$. We start with a state which is symmetric on all the basis states, and use this vertex-dependent coin to ``accumulate" probability in the states associated with $x$. Note that the number of coin states associated with each vertex is $d=4$, and $S$ and $F$ are both Hermitian; hence $(SF)^{-1} = FS$.

\paragraph*{Notation:} We have already defined $\ket{S_c}$ as the uniform superposition of all $d=4$ coin basis states.  We similarly define
\begin{equation}
\ket{S_v}=\frac{1}{\sqrt{N}}\sum_v\ket{v}
\end{equation}
as the uniform superposition of all vertex states.  We denote $\ket{S_c, S_v} = \ket{S_c}\otimes \ket{S_v}$.

Consider the continuous limit of the algorithm with Hamiltonian $S-F$, as described in the previous section. The probability of being at a vertex $v$ in the continuous-time walk is the probability of being in the subspace spanned by the vectors associated with $v$ in the DTQW.

\begin{thm}
\label{thm:search}
The continuous limit of the DTQW search algorithm described in \cite{AAmbainis08} is a CTQW search algorithm with Hamiltonian $S-F$, with the same time complexity (that is, $O(\sqrt{N}\log{N})$).
\begin{proof}
The initial state is $\ket{S_c,S_v}$.  Let $x$ be the marked vertex. The theorem follows from five partial results:
\begin{claim}
\label{claim:first}
$(S-F)\ket{S_c,S_v} = \frac{2}{\sqrt{N}}\ket{S_c,x}$
\begin{proof}
$S\ket{S_c,S_v} = \ket{S_c,S_v}$, $F\ket{S_c,S_v} = \ket{S_c,S_v} -2\braket{x\mid S_v}\ket{S_c,x}$, and $\braket{x\mid S_v} = {1}/{\sqrt{N}}$. Putting these together yields the result.
\end{proof}
\end{claim}

\begin{claim}
\label{claim:second}
If $\ket{j}$ is an eigenvector of $SF$ with eigenvalue $e^{i\phi_j}$ then
\[
\ket{j}\braket{j \mid S_c,S_v}
= -i\frac{e^{i{\phi_j}/{2}}}{\sqrt{N}\sin({\phi_j}/{2})}\ket{j}\braket{j \mid S_c,x} .
\]
\begin{proof}
\begin{eqnarray*}
\braket{j \mid S_c,S_v} &=& e^{-i\phi_j}\bra{j}FS\ket{S_c,S_v} \\
&=& e^{-i\phi_j}\bra{j}( \ket{S_c,S_v} -2\braket{x\mid S_v}\ket{S_c,x}) .
\end{eqnarray*}
Solving for $\braket{j \mid S_c,S_v}$ yields the result.
\end{proof}
\end{claim}

\begin{claim}
\label{claim:time-evo}
\begin{eqnarray}
\ket{\Psi(t)} &=& e^{-i(S-F)t}\ket{S_c,S_v} \\
&=& \sum_j  \left(\cos(\lambda_jt) + e^{-i\frac{\phi_j}{2}} \sin(\lambda_jt)\right)
  \ket{j}\braket{j\mid S_c,S_v} . \nonumber
\end{eqnarray}
\begin{proof}
Follows from Lemma ~\ref{lem:exp}, Claim ~\ref{claim:first} and Claim ~\ref{claim:second}.
\end{proof}
\end{claim}

\begin{claim}
\label{claim:probab}
The probability of being at the node $x$ at a time instant $t$ is $|\braket{S_c,x\mid\Psi(t)}|^2$.
\begin{proof}
The initial state is symmetric with respect to all directions about the marked node $x$.  The unitaries $S$ and $F$ preserve this property.  This means that the amplitude of being at each coin state associated with $x$ at any time $t$ is the same. The result follows from this. 
\end{proof}
\end{claim}

The initial state is $\ket{\Psi(0)} = \ket{\Psi_0} = \ket{S_c,S_v}$. 
In \cite{AAmbainis08} they show that there are eigenstates $\ket{w_{\alpha}}$ and $\ket{w_{-\alpha}}$ of $U^{\prime} =SF$  with eigenvalues $e^{i\alpha}$ and $e^{-i\alpha}$, respectively, such that
\begin{equation}
\ket{S_c,S_v} = \frac{1}{\sqrt{2}} (\ket{w_{\alpha}}-\ket{w_{-\alpha}})+\ket{\Phi_{rem}}
\end{equation}
and 
\begin{equation}
\|\braket{S_c,x\mid w_{\alpha}}+\braket{S_c,x\mid w_{-\alpha}}\| =  \Theta(\frac{1}{\sqrt{\log{N}}}) ,
\end{equation}
where $\|{\ket{\Phi_{rem}}}\| = \Theta (\frac{1}{\log{N}})$ and $\alpha = \Theta(\frac{1}{\sqrt{N}})$.  From Lemma~\ref{lem:exp}, it follows that
\begin{multline}
\ket{\Psi(t)} =[f(\alpha,t)\ket{w_{\alpha}}\braket{w_{\alpha}\mid S_c,S_v}] \\ 
= [f(-\alpha,t)\ket{w_{-\alpha}}\braket{w_{-\alpha}\mid S_c,S_v}] + \ket{\Phi_{rem1}(t)} .
\label{search_vector}
\end{multline}
Here, $f(a,t) = \cos(\theta_{a}t)+e^{-i\frac{a}{2}}\sin(\theta_{a}t)$,  $\ket{\Phi_{rem1}(t)}$ is a vector perpendicular to both $\ket{w_{\alpha}}$ and $\ket{w_{-\alpha}}$, and $\theta_{a} = 2\sin(\frac{a}{2})$.

\begin{claim}
\label{claim:bound_rem}
The magnitudes in Eq.~(\ref{search_vector}) are:
\[
\|\ket{\Phi_{rem_1}(t)}\| = O\left(\frac{1}{\log{N}}\right) ,
\]
\[
\|\braket{Sc,Sv\mid w_{\pm\alpha}}\mp{\sqrt{1/2}}\| = O\left(\frac{1}{\log{N}}\right) .
\]
\begin{proof}
Taking the state at $t=0$, 
\begin{multline}
\ket{\Psi(0)} = \ket{S_c,S_v} = \frac{1}{\sqrt{2}}(\ket{w_\alpha} -\ket{ w_{-\alpha}})+\ket{\Phi_{rem}} \\
=\braket{Sc,Sv\mid w_{\alpha}}\ket{w_{\alpha}}+\braket{Sc,Sv\mid w_{-\alpha}}\ket{w_{-\alpha}} + \ket{\Phi_{rem1}(0)} ,
\end{multline}
which implies
\begin{multline}
\ket{\Phi_{rem}} = \left( \braket{Sc,Sv\mid w_{\alpha}}-{\sqrt{1/2}}\right) \ket{w_{\alpha}} +\\
\left(\braket{Sc,Sv\mid w_{-\alpha}}+{\sqrt{1/2}}\right)\ket{w_{-\alpha}} +\ket{\Phi_{rem1}} .
\end{multline}
Since all three vectors on the RHS are orthogonal, their norms must each be less than or equal to the norm of $\ket{\Phi_{rem}}$ on the LHS.  Since $\|\ket{\Phi_{rem}}\| = \Theta({1}/{\log{N}})$, the result follows.
\end{proof}
\end{claim}

\begin{claim}
At $t={\pi}/{2\theta_{\alpha}}$,
\[
\|\braket{S_c,x\mid\Psi(t)}\| = \Omega\left(\frac{1}{\sqrt{\log{N}}}\right) .
\]
\begin{proof}
\begin{multline}
\ket{\Psi({\pi}/{2\theta_{\alpha}})} = e^{-i{\alpha}/{2}}\ket{w_{\alpha}}\braket{w_{\alpha}\mid S_c,S_v} \\-  e^{i{\alpha}/{2}}\ket{w_{-\alpha}}\braket{w_{-\alpha}\mid S_c,S_v} + \ket{\Phi_{rem1}({\pi}/{2\theta_{\alpha}})} .
\end{multline}
The result follows from Claim ~\ref{claim:bound_rem}.
\end{proof}
\end{claim}

With this last result we can prove the theorem.  The probability of the particle being at the node $x$ is $\Omega({1}/{\log N})$ at $t ={\pi}/{2\theta_{\alpha}}$ where $\theta_{\alpha} = \Theta({1}/{\sqrt{N}})$. Repeating the algorithm $O(\log{N})$ times gives us a constant probability of finding the marked item. The time complexity of the continuous time search algorithm is therefore $O(\sqrt{N}\log{N})$.
\end{proof}
\end{thm}

This result is in contrast to the proof in \cite{Childs2004} that any CTQW search algorithm over a $\sqrt{N}\times\sqrt{N}$ grid with circular boundary conditions takes $\Omega(N)$ time. But, by defining the CTQW search as the continuous time limit on the related graph, as described in Sec.~\ref{sec:limit}---that is, a {\it coined} CTQW---the search becomes as efficient as the DTQW search algorithm.

\section{Effect of graph symmetries}

In \cite{HariKrovi2007} it was shown that if a DTQW with an appropriate unitary evolution is defined on a graph with symmetries, then for certain symmetric initial conditions the walk can be reduced to a walk on its {\it quotient graph}.  This is a smaller graph obtained by identifying certain groups of vertices and edges. We briefly review the reduction.

Let $H$ be a subgroup of the symmetry group over $n$ letters, where $n$ is the dimension of the state-space of a DTQW over $G$. The elements of $H$ are such that $\forall h \in H$, $[\sigma(h),S]=0$, where $\sigma(h)$ is the matrix representation of the permutation and $S$ is the shift operator. It is shown in \cite{HariKrovi2007} that if $U = SF$ (where $F$ is the coin flip operator) and $[U,\sigma(h)]=0$ for every $h\in H$, then there exists a common set of  eigenvectors $A = \set{\ket{\mathcal{O}_x}}$, with eigenvalue $1$ for all $\sigma(h)$, such that
\begin{equation}
\forall \ket{\mathcal{O}_x} \in A, \ U\ket{\mathcal{O}_x} \in A .
\end{equation}
Locality of the graph is preserved, in the sense that two vectors $\ket{\mathcal{O}_x}$ and $\ket{\mathcal{O}_y}$  in $A$ are connected if and only if every component $\ket{i}$ of $\ket{\mathcal{O}_x}$ is connected to some component $\ket{j}$ of $\ket{\mathcal{O}_y}$ in the graph $G$.  Mathematically, this means that $\bra{\mathcal{O}_y}S\ket{\mathcal{O}_x} \neq 0$ if and only if for every state  $\ket{i}$ such that $\braket{i\mid\mathcal{O}_x} \neq 0$ there exists a state $\ket{j}$ such that $\braket{j\mid\mathcal{O}_y} \neq 0$  and $\bra{j}S\ket{i} \neq 0$. 

The ``quotient graph" $G_H$ is the graph whose states are the vectors in $A$ and whose connectivity is defined as above.  Two states $\ket{\mathcal{O}_x}$ and $\ket{\mathcal{O}_y}$ are associated with the same vertex in $G_H$ if, for every state $\ket{c_1,v}$ of $G$ such that $\braket{c_1,v\mid\mathcal{O}_x} \neq  0$, there exists a coin state labeled by $c_2$ such that $\braket{c_2,v\mid\mathcal{O}_y} \neq  0$.

We can easily see that if $[U,\sigma(h)]=0$ and $[S,\sigma(h)] = 0$, then $[F,\sigma(h)] = 0$.  Let $F$ be Hermitian. Then by the results of Section ~\ref{sec:limit} the continuous limit of the DTQW is generated by the Hamiltonian $\mathcal{H} = S+ F$ over the related graph $G^{\prime}$. From the above, $[\sigma(h),\mathcal{H}]=0$.

\begin{thm}
$\mathcal{H}$ defines a CTQW over $G^{\prime}_H$ whose basis states are the basis states of $G_H$.
\begin{proof}
The proof is similar to as given in \cite{HariKrovi2007}. Since the $\{\ket{\mathcal{O}_x}\}$ are eigenvectors of $\sigma(h)$ with eigenvalue $1$,
\begin{equation}
\sigma(h)\mathcal{H}\ket{\mathcal{O}_x} = \mathcal{H}\sigma(h)\ket{\mathcal{O}_x} = \mathcal{H}\ket{\mathcal{O}_x} .
\end{equation}
This proves that $\mathcal{H}$ is an operator on the space spanned by the vectors in $A$. As in Section ~\ref{sec:limit}, we can define the graph $G^{\prime}_H$ whose vertices correspond to the vectors in $A$. Hence, $\mathcal{H}$ defines a CTQW over $G^{\prime}_H$, or a {\it coined} CTQW over $G_H$.
\end{proof}
\end{thm}

It straightforwardly follows that this walk is the continuous-time limit of a DTQW over the quotient graph $G_H$.

\begin{thm}
The family of unitary operators $U(s) = e^{-i\frac{\pi}{2}s(S-I)}e^{-i\frac{\pi}{2}s(F-I)}$, parametrized by $s \in (0,1]$, define a local transformation over $G_H$, which for $s=1$ is a DTQW.  Hence, the CTQW given by the Hamiltonian $\mathcal{H}$ over $G^{\prime}_H$ can be seen as the continuous limit of a DTQW over the quotient graph in the limit $s \rightarrow 0$.
\begin{proof}
By the proof of  Lemma ~\ref{lem:local},  $U(s) =  a^2I + ab(S+F)+ b^2(SF)$. $I$, $S+F$ and $SF$ are local transformations over $G_H$. Hence $U(s)$ is a local transformation over $G_H$ $\forall$ $s \in [0,1]$.
\[
U(s) = I - i\frac{\pi}{2}s(S+F-2I) + O(s^2) .
\]
Taking the limit $s \rightarrow 0$, as shown in Section ~\ref{sec:limit}, we see that $\mathcal{H} = S+F$ is a Hamiltonian that defines a CTQW over $G^{\prime}_H$.
\end{proof}
\end{thm}

\section{Conclusion}

In classical random walks, there is a straightforward sense in which continuous time random walks are a limit of discrete time random walks.  Both can be defined on the same graph, with behaviors that are opposite limits of a continuous family of evolution rules.  Because discrete-time and continuous-time quantum walks are defined on state spaces with different dimensions, constructing such a correspondence is not simple.  A small number of attempts have been made to overcome this problem.

This paper presents a different approach.  For a particular class of DTQWs with Hermitian coins and shift operators, the continuous-time limit of a DTQW on a graph $G$ is a continuous-time walk on a different, but related, graph $G'$.  The two evolution rules can be defined as opposite limits of a continuous family of evolution rules, just as in the classical case.  We can think of this walk on the graph $G'$ as being a {\it coined} continuous time quantum walk.

Because the continuous-time limit is defined on the same space as the DTQW, it shares many properties with the original walk.  We have shown, for example, that the continuous-time limit of the DTQW search algorithm has the same $\sqrt{N}$ speed-up as the original algorithm; the usual CTQW on the same graph has no speed-up.

Similarly, DTQWs on symmetric graphs can exhibit a reduction to a DTQW on a smaller quotient graph.  This property is closely connected to the existence of quantum speed-ups in certain quantum-walk based algorithms \cite{ChildsCleve2003}.  We have shown the the CTQW limit of this walk shares this reduction to a walk on the quotient graph.

The ability to take such limits---and the existence of coined CTQWs---adds another tool to the arsenal of quantum walks, and one that deserves to be further explored.  Moreover, the interesting question of why the use of coined walks can sometimes produce speed-ups also deserves further study.  In addition to their own beautiful properties, quantum walks have proven to be a fertile field for the study of quantum algorithms.  We hope to illuminate these questions in future work.

\bibliographystyle{apsrev4-1}

\bibliography{bib2}
\end{document}